\documentclass[creativecommons]{eptcs}
\providecommand{\event}{LoCoCo 2011} \usepackage{breakurl}             
\usepackage{url}
\usepackage{alltt}
\title{Aligning component upgrades\footnote{Partially supported by the European Community's
    7th Framework Programme (FP7/2007-2013), grant agreement
    n${}^\circ$214898, ``Mancoosi'' project. Research partially performed at IRILL.}}
\author{Roberto Di Cosmo
        \institute{Univ Paris Diderot, Sorbonne Paris Cit\'e,\\ Laboratoire PPS, UMR 7126,\\ INRIA Rocquencourt\\ F-75205 Paris France} 
        \email{roberto@dicosmo.org}
      \and 
        Olivier Lhomme\institute{IBM France,\\ 1681, route des Dolines,\\ 06560 Sophia Antipolis, France} 
	\email{om.lhomme@gmail.com, olivier.lhomme@fr.ibm.com} 
      \and 
        Claude Michel
        \institute{I3S (UNS-CNRS),\\ 2000 route des Lucioles, BP 121,\\06903 Sophia Antipolis Cedex, France} 
        \email{Claude.Michel@i3s.unice.fr}
}

\def\titlerunning{Aligning component upgrades}
\def\authorrunning{R. Di Cosmo, O. Lhomme \& C. Michel}
\long\def\ignore#1{\relax}

\begin{document}
\newtheorem{definition}{Definition}
\maketitle

\begin{abstract}
Modern software systems, like GNU/Linux distributions or Eclipse-based
development environment, are often deployed by selecting components out of large
component repositories.  Maintaining such software systems by performing
component upgrades is a complex task, and the users need to have an expressive
preferences language at their disposal to specify the kind of upgrades they are
interested in. Recent research has shown that it is possible to develop solvers
that handle preferences expressed as a combination of a few basic criteria used
in the MISC competition, ranging from the number of new components to the
freshness of the final configuration.  In this work we introduce a set of new
criteria that allow the users to specify their preferences for solutions with
components aligned to the same upstream sources, provide an efficient encoding
and report on the experimental results that prove that optimising these
alignment criteria is a tractable problem in practice.
\end{abstract}

\section{Introduction}

Recent research, in part fostered by the Mancoosi project\footnote{\url{http://www.mancoosi.org}},
has focused on the complex problem of handling upgrades in component based software systems,
with a particular attention to the case of GNU/Linux distributions, which contain several
tens of thousands of components. Installing components (called packages in the world of
distributions) may be complex: each component may need some extra components to be installed, as
described in its metadata by \emph{dependencies}, and may be incompatible with some other ones, as described
in its metadata by \emph{conflicts}. Indeed, determining whether a component can
be installed is NP-complete ~\cite{EdosAse06}, but problem instances arising in practice turn
out to be tractable by modern solvers~\cite{EdosAse06,tucker-opium,LeBerreParrain2008}.
These practical results opened the way to explore not just the question of finding a way of
installing some components, but the \emph{best} way of doing so, according to some criteria
that capture the user preferences and needs.\\

The Mancoosi International Solver Competition (MISC)\footnote{\url{http://www.mancoosi.org/misc}} was 
established with the goal to distill interesting problems from real-world GNU/Linux distribution
upgrade scenarii, and present them to the solver research community. The problems are encoded
in documents written using a common format, CUDF~\cite{mancoosi-tr3}, that describe the universe
of available components with their interdependencies and the user request; the solvers are 
requested to find solutions that are ranked according to user preferences which are 
currently built by composing a few basic criteria using aggregation functions like the lexicographic
ordering.

There are five basic criteria currently used in MISC: \emph{removed},
\emph{changed}, \emph{new}, \emph{notuptodate} and \emph{unsatrecommends}, which
capture intuitive properties of a solution to an upgrade problems, like the
number of removed components or the number of components that are not the most
up to date. They are summarised in Table~\ref{tab:optcriteria} where $I$ is the
initial installation and $S$ is a proposed new installation. We write
$V_p(X,\mathit{name})$ for the set of versions in which \emph{name} (the name of a
component) is installed in $X$, where $X$ may be $I$ or $S$. That set may be
empty (\emph{name} is not installed), contain one element (\emph{name} is
installed in exactly that version), or even contain multiple elements in case a
component is installed in multiple versions.
These criteria and aggregation functions are an important starting point for this research, 
but are not sufficient to capture all the important properties of an upgrade : identifying
new basic criteria and new aggregation function is an important activity, that will help
improve the algorithms and tools available for maintaining the complex software systems
of tomorrow.

\begin{table}[t]
  \caption{Optimization criteria\strut}
  \label{tab:optcriteria}
  \begin{tabular}{l@{}l@{}l}
    \hline
    $removed(I,S)$ & =
      & \{$name ~|~ V_p(I,name) \neq \emptyset \mbox{ and } V_p(S,name) = \emptyset$\} \\
    $new(I,S)$ & =
      & \{$name ~|~ V_p(I,name) = \emptyset \mbox{ and } V_p(S,name) \neq \emptyset$\} \\
    $changed(I,S)$ & = & \{$name ~|~ V_p(I,name) \neq V_p(S,name)$\} \\
    $notuptodate(I,S)$ & =
      & \{$name ~|~ V_p(S,name) \neq \emptyset$ 
      $\mbox{ and does not contain the most recent version of name in } S$\}  \\
    $unsatrec(I,S)$ & =
      & \{$(name,v,c)$ | $v$ is an element of $V_p(S,name)$ 
        $ \mbox{ and } (name,v) \mbox{ recommends } ..., c, ...$ \\
     & & $ \mbox{ and } c \mbox{ is not satisfied by } S$\} \\
     \hline
  \end{tabular}
\end{table}

\paragraph*{Contribution}

In the context of complex software systems, we can expect that configurations
containing synchronized components be more robust, for multiple reasons:
synchronised components have been in general developed together, and more
thoroughly tested. Component metadata may contain, or can be enriched with,
information about synchronization (for example, via a \emph{Source version}
field), that can be exploited to search for synchronised configuration.

In this paper, we present a new criterion, \emph{component alignment}, which
measures the synchronization of closely related components in an
installation and which is not expressible using the existing criteria
used in the MISC competition. Then, we show how to encode it using current
solver technology, and present experimental results that show that it
is tractable in practice.

\section{Component alignment}

In complex software systems, like GNU/Linux distributions, components do not exist in isolation,
but are very often related to each other, even if they may be installed independently: the
documentation of a program, for example, is not necessary to run it, but they are both present,
the user expects them to be of the same version, or, in other terms, to be \emph{aligned}.\\

With the current basic preferences used in MISC, it is not possible to express
this alignment property, and one can see that even the best solutions in the
MISC 2010 competition may contain strange component combinations : for example,
the solutions found by the competition entrants for the problem 
eeee44ce\footnote{See \url{http://data.mancoosi.org/misc2010/results/problems/debian-dudf/eeee44ce-5407-11df-b11f-00163e7a6f5e.cudf.bz2}}
in the trendy track, contains surprising combinations like

\begin{center}
\begin{tabular}{ll}
Package & version\\\hline
  aptitude-doc-fr & 0.6.1.5-3\\
  aptitude & 0.4.11.11-1+b2\\
\end{tabular}
\end{center}

\noindent or even \\
 
 \begin{center}
\begin{tabular}{ll}
Package & version\\\hline
  linux-libc-dev & 2.6.32-9\\
  linux-source-2.6.30 & 2.6.30-8squeeze1\\
 \end{tabular}
 \end{center}

These are potential sources of confusion for a user that finds documentation way ahead of the
installed binaries or sources way behind the installed libraries. In the case of mixed versions
of important libraries, like \texttt{gs} in version \texttt{8.64~dfsg-1+squeeze1}
which is used with \texttt{gs-common} version \texttt{8.71~dfsg-4} in the same example, one
can even experience real incompatibilities, due to combination of components that have not
been thoroughly tested together.\\

Obtaining an aligned installation by hand is quite painful, because of the
number of involved packages: it is really necessary to be able to express the
preference concisely via a criterion.
To help the users that want to avoid these inconsistencies, we propose to exploit
the information about the package \emph{source}, which is present in the metadata 
of mainstream distributions.

In Debian, for example, packages which are built from the same source package
carry in their metadata two pieces of information: 
\begin{itemize}
 \item a \emph{source} property,that specifies the name of the source package; 
       for example, both packages related to the Linux kernel in the example above
       have a \emph{source} property with the same value \texttt{linux-2.6})
 \item the version of the source used to build them; this information is 
       encoded in the CUDF documents coming from Debian distributions in a \emph{sourceversion}
       property; for example, the two packages related to the Linux kernel in the example above
       are built from two different versions, \texttt{2.6.32-9} and \texttt{2.6.30-8squeeze1},
       of the same source \texttt{linux-2.6}.
\end{itemize}
     
Using this information, one can define what it means for an installation to be \emph{aligned}.

\begin{definition}[Alignment]
An installation $I$ is \emph{source aligned} if all installed packages built from a same source $s$ are
actually built from the same version of this source.
\end{definition}

In other terms, $I$ is aligned if all packages $p_i$ having the same value for the \emph{source} property
also have the same value of the \emph{sourceversion} property.

We remark here that \emph{the version of a package}, and \emph{the version of the source} from which they are built
do not necessarily coincide, and packages built from the same version of the same source may carry different
package versions, so that using the \emph{version of the source} as an alignment criterion is the best way of
knowing whether a set of packages is aligned, without the need to guess similarity of packages by inspecting
their package versions.

\paragraph*{Notations}

In the following, we write $\cal S$ for the set of sources of the problem to solve.

We note $\{p_i\}_{i=1..n}$ the set of all available packages. For simplifying 
the notation, $p_i$ will also denote a 0-1 variable that expresses that
package $p_i$ is installed; when the context is not enough to resolve
the ambiguity we write package $p_i$ or variable $p_i$.

The relationships between the sources, the packages and their versions
will be expressed with the following functions:
\begin{itemize}
\item $V(s)$ denotes the set of versions of source $s \in \cal S$; 
\item $V(p)$ denotes the version of the source of package $p$;
\item $P(s,v)$ denotes the set of packages belonging to version $v$ of source $s$;
\item $S(p)$ denotes the value of the source property of package $p$.
\end{itemize}
For example, $p \in P(s,v) \Leftrightarrow V(p)=v$ and  $S(p)=s$.
\section{Measuring  \emph{unalignment}}

In order to choose among different possible installations, we need to be able to
measure how far we are from an aligned solution; for this, we need a measure of
\emph{unalignment} of a solution to a user query, that can be then used as an
objective function to minimize.

It turns out that there are quite a few different ways of defining such a
notion, with varying cost and expressiveness.  We discuss them in the following
sections, where we present the different possible definitions.
An encoding for MIP solvers, along the lines of~\cite{DBLP:journals/corr/abs-1007-1020}, 
is given in detail in Section~\ref{s:milp}.

\subsection{Counting unaligned packages}
A first approach to building a measure of unalignment is to count the number of packages $p_i$ which
are installed and not source aligned. This can be expressed formally as the cardinality of a set:

\[
 unaligned_p = card \{p_i | i\in [1..n], p_i=1, \exists j p_j=1, S(p_j)=S(p_i),  V(p_j)\neq V(p_i) \}
\]

Note that in our notation, variables $p_i$ and $p_j$ are equal to 1 mean that packages $p_i$ and
$p_j$ are installed. The above set contains all packages that are installed
and such that another package with the same source in another version
is also installed.

To obtain an installation that is as aligned as possible, it is then
enough to minimize \(unaligned_p\), the cardinality of the set.

\subsection{Counting (sorted) unaligned package pairs}

A second approach is to count the number of \emph{pairs of packages} $(p_i,p_j)$ which are both installed and not aligned. This can
be done by computing the cardinality of a slightly different set:

\[
 unaligned_{pp} = card \{ (p_i,p_j) | i,j\in [1..n], i<j, p_i=1, p_j=1, S(p_j)=S(p_i),  V(p_j)\neq V(p_i) \}
\]

The interest of this approach is to be much more discriminating than
the \(unaligned_p\) criteria (see Section~\ref{discussCrit}).
Nevertheless, a drawback may be that, as it implicitly weights a
cluster up to the square of its size, a small qualitative improvement
of a large and very unaligned cluster may strongly dominate clear
qualitative improvements of some other smaller or almost aligned
clusters.

\subsection{Counting version changes}
In this third approach, the size of the cluster is not as important as in the \(unaligned_{pp}\) criteria: it counts the number
of version changes in a cluster. For example, consider a cluster with six installed packages that involve three different
source versions: there will be two version changes. Formally:

\[
 unaligned_{vc}= \sum_{s \in \cal S} max(0,numberOfVersions(s)-1)
\]
where:
\[
 numberOfVersions(s) = card \{ V(p_i) | i\in [1..n], p_i=1, S(p_i)=s\} 
\]
Note that \(numberOfVersions(s)\) is the number of installed versions
of the source $s$; thus, when this number is greater than 0, we need
to subtract 1 to get the number of version changes.

\subsection{Counting unaligned source clusters}

Finally, one can use a much coarser granularity, counting only the \emph{source clusters} which are unaligned,
independently of the number of pointwise unalignments among packages of the same cluster, by using

\[
 unaligned_c =  card \{s | {s \in \cal S},  \exists i\in [1..n], \exists j\in [1..n], p_i=1, p_j=1, S(p_j)=S(p_i)=s,  V(p_j)\neq V(p_i) \}
\]

\subsection{Discussion of the different alignment criteria}\label{discussCrit}
The different alignment criteria differ by their weighting policies.
The number of unaligned source clusters $unaligned_c$ and the number
of unaligned packages $unaligned_p$ are very close, except that the
criterion $unaligned_c$ does not take into account the size of the
clusters, whereas the criterion $unaligned_p$ weights a cluster by its
size (each time a cluster of size $k$ is unaligned, $k$ packages are
unaligned).  The criterion $unaligned_{pp}$ is more discriminating by
weighting a cluster by its pairwise unalignment, which may be really
interesting, but it makes the implicit assumption that packages of a
cluster are totally interdependant. When this assumption is too strong
and the size of the cluster is large, the weight of a $k$-sized
cluster can be as large as $k^2$, and alignments in large clusters may
dominates too strongly alignments in small clusters.  The criterion
$unaligned_{vc}$, based on version changes, provides an interesting
intermediate solution: the weight of the cluster is the number of
different versions in that cluster.

To see in practice what each of the above criterion actually captures, it is
useful to compare the results on a simple example. Let's consider a cluster
$c=\{p_1,p_2,p_3,p_4\}$ comprising 4 packages of the same source, with package
versions among $1,2,3,4$, and a few possible unaligned configurations.\\

\begin{center}\small
\begin{tabular}{|c|r|r|r|r|}
\hline
version       & unaligned & unaligned & unaligned & unaligned \\
configuration & packages & pairs & version changes & clusters \\
\hline
1,1,1,1 &0&0&0&0\\
1,1,2,1 &4&3&1&1\\
1,1,2,2 &4&4&1&1\\
1,1,2,3 &4&5&2&1\\
1,2,3,4 &4&6&3&1\\
\hline
\end{tabular}
\end{center}

\section{Efficiently encoding the criteria using MIP}\label{s:milp}

This section describes an integer programming encoding of the
unaligned criteria presented above. It is particularly efficient in
practice with a MIP solver. Note that a clausal form of these criteria
can also be obtained for using a SAT solver (see the Appendix).\\

As a first step, the problem is reduced to the subset of sources with more than one
source version.

\subsection{packages}

The number of unaligned packages is computed using the following formulae
$$nu_{packages} = \sum_{p_j \in P(s,v), v \in V(s), s \in {\cal S}} nu_{p_j}$$
where $nu_{p_j}$ is a binary variable whose value is one if $p_j$ is installed
and not aligned and zero otherwise.

Each  $nu_{p_j}$ is handled by the following set of constraints
$$nu_{p_j} \leq p_j$$
which forces $nu_{p_j}$ to $0$ if package $p_j$ is not installed, and
$$nu_{p_j} \leq \sum_{v \in V(S(p_j)), v \neq V(p_j)}i_{s,v}$$
where $s = S(p_j)$ and $i_{s,v}$ is binary variable whose value is $1$ if
any package of version $v$ from source $s$ is installed and zero otherwise.
Therefore, the previous constraint forces $nu_{p_j}$ to zero if none of the other
versions of source $s$, different from $V(p_j)$, has an installed package.
$nu_{p_j}$ is also involved in the following set of constraints
$$\forall v \in V(S(p_j)),\,\, v \neq V(p_j),\,\, nu_{p_j} + 1 \geq p_j + i_{s,v}$$
which ensures that if $p_j$ is installed and one of the versions of $s$ different from the source version
of $p_j$ has an installed package, then $nu_{p_j}$ is set to one.

Finally, constraints are added to handle the $i_{s,v}$ variables.
The first constraint ensures that $i_{s,v}$ gets the value zero if none of the 
packages of version $v$ from source $s$ is installed
$$i_{s,v} \leq \sum_{p_j \in P(s,v)} p_j$$
The second set of constraints sets $i_{s,v}$ to 1 whenever at least one of the 
packages of version $v$ from source $s$ is installed
$$\forall p_j \in P(s,v),\,\, p_j \leq i_{s,v}$$
Note that variables $i_{s,v}$ are also used in the encoding of the two last unaligned criteria.

\subsection{pairs}

The number of unaligned pairs 
$$nu_{pairs} = \sum_{p_j \in P(s,v), v \in V(s), s \in {\cal S}, p_k \in P(s,v'), v' \in V(s), v' \neq v} u_{p_j,p_k}$$
where each $u_{p_j,p_k}$ is subject to the three following constraints:
$$u_{p_j,p_k} \leq p_j  \,\,\wedge\,\, u_{p_j,p_k} \leq p_k  \,\,\wedge\,\, u_{p_j,p_k} + 1 \geq p_j + p_k$$
The two first constraints insure that $u_{p_j,p_k} = 0$ if either $p_j$ or $p_k$ is not installed.
Last constraint sets $u_{p_j,p_k}$ iff both  $p_j$ and $p_k$ are installed.

\subsection{version changes}

The number $nu_{vc}$ of version changes is given by the following formulae:
$$nu_{vc} = \sum_{s\, \in\, {\cal S}} nc_s$$
where each $nc_s$ is subject to
$$nc_s = nb_{inst,s} - \delta_s$$
where each $\delta_s$ is subject to 
$$|V(s)| * \delta_s \geq nb_{inst,s} \,\,\wedge\,\, nb_{inst,s} \geq \delta_s$$
The first constraint sets $\delta_s$ to $1$ iff $nb_{inst,s} \geq 1$, and
the second one sets $\delta_s$ to $0$ iff $nb_{inst,s} = 0$.
The  $nb_{inst,s}$ variable simply sum up the number of installed source versions (i.e., the number
of source versions with at least one installed package). Thus,
$$nb_{inst,s} = \sum_{v \in V(s)} i_{s,v}$$

\subsection{clusters}

The number of unaligned clusters of source is given by the following formulae:
$$nu_{clusters} = \sum_{s\, \in\, {\cal S}}u_s$$
where each $u_s$ is subject to
$$|V(s)| * u_s + 1 \geq nb_{inst,s} \,\,\wedge\,\,nb_{inst,s} \geq 2 * u_s$$
The first constraint sets $u_s$ to $1$ iff $nb_{inst,s} \geq 2$, while the second one
forces $u_s$ to $0$ iff  $nb_{inst,s} \leq 1$.
$nb_{inst,s}$ has the same definition as in the unaligned version changes.

\nocite{Predictions2011,MancoosiHotSwUp2008,aptpbo,Lococo2010-Argelich,ArgelichIJCAI09,mpm2011,LeBerreParrain2008}

\section{Experimental validation}

We implemented the four alignment criteria introduced above in an experimental
branch of the \texttt{mccs}
tool\footnote{\url{http://users.polytech.unice.fr/~cpjm/misc/mccs.html}}, which
uses MIP instead of the Boolean encodings, and includes several optimizations
with respect to the simple encodings detailed above.

We have run the solver on the Debian category of the problems of the MISC-2010 competition
and of the 4th run of the Misc Live competition
with a realistic optimization function that requires, in lexicographic order to first minimize
removal, and then minimize unalignment, using each of the four different criteria for unalignment.
The results\footnote{Though the two sets of Debian problems share some problem IDs, there are different
problems as testified by the problem sizes and the different times.} of running this experiments on an Intel Core I7-2720QM at 2.20GHz are
given in Figure~\ref{fig:experiments} for the MISC-2010 competition and
 in Figure~\ref{fig:experiments2} for the 4th run of the Misc Live competition.
In these tables, the \emph{size} column gives respectively, the number of sources (with more than one version) of
the problem, the total amount of versions, the total amount of packages (corresponding to the selected sources/versions), and,
the number of unique pairs. 
The \emph{removed} column gives the time (in seconds) required to optimise the problem according to the sole removed criterion,
as well as, in brackets, the number of unaligned packages, pairs, version changes and clusters of the solution.
Last four columns give the amount of time required to solve the problem minimizing removal and the chosen unalignment,
as well as, in brackets, the number of unalignments.
Note that, for the sake of fairness, CPLEX, the underlying MIP solver, has been limited to one thread.

\begin{figure}[t]
\begin{center}\scriptsize
\begin{tabular}{lllrrrr}
MISC problem id & size (\#srcs,\#vs,\#pkgs,\#pairs) & removed & packages & pairs & version changes & clusters\\
\hline
103c9978 & 183,531,531,377 & 0.71 (25,15,10,10) & 1.10 (2) & 1.04 (1) & 1.18 (1) & 1.07 (1) \\
1dcce248 & 3833,12319,12319,15595 & 5.00 (0,0,0,0) & 11.56 (0) & 11.24 (0) & 13.55 (0) & 11.14 (0) \\
218091ce & 3833,12319,12319,15595 & 4.62 (0,0,0,0) & 10.54 (0) & 10.00 (0) & 12.20 (0) & 10.07 (0) \\
29180036 & 183,531,531,377 & 0.69 (25,15,10,10) & 1.08 (2) & 1.05 (1) & 1.16 (1) & 1.08 (1) \\
2f690324 & 3834,12301,12301,15558 & 4.52 (0,0,0,0) & 10.07 (0) & 9.85 (0) & 11.88 (0) & 9.87 (0) \\
3e4f8550 & 3805,12033,12033,14766 & 4.71 (0,0,0,0) & 10.73 (0) & 10.44 (0) & 12.00 (0) & 10.63 (0) \\
412959c6 & 552,1193,1193,736 & 0.88 (0,0,0,0) & 1.47 (0) & 1.38 (0) & 1.54 (0) & 1.32 (0) \\
56ae4afa & 3805,12033,12033,14766 & 4.86 (0,0,0,0) & 10.49 (0) & 9.78 (0) & 12.36 (0) & 10.38 (0) \\
58a4a468 & 3857,12136,12136,14801 & 4.71 (0,0,0,0) & 8.54 (0) & 8.17 (0) & 15.91 (0) & 7.95 (0) \\
688250e8 & 3833,12319,12319,15595 & 7.63 (0,0,0,0) & 14.97 (0) & 12.33 (0) & 17.80 (0) & 14.56 (0) \\
7266f636 & 3840,12371,12371,15750 & 4.99 (0,0,0,0) & 11.45 (0) & 11.25 (0) & 13.24 (0) & 10.93 (0) \\
7e7e0b16 & 3857,12136,12136,14801 & 4.78 (0,0,0,0) & 8.53 (0) & 8.17 (0) & 16.00 (0) & 8.22 (0) \\
8ad21cec & 3840,12371,12371,15750 & 5.00 (0,0,0,0) & 11.45 (0) & 11.24 (0) & 13.33 (0) & 11.06 (0) \\
9bb87ab4 & 3833,12319,12319,15595 & 4.66 (0,0,0,0) & 10.66 (0) & 10.08 (0) & 12.39 (0) & 10.17 (0) \\
cb0e73b0 & 918,1940,1940,1127 & 1.12 (0,0,0,0) & 1.84 (0) & 1.72 (0) & 2.87 (0) & 1.80 (0) \\
e0bd67a6 & 124,354,354,259 & 0.62 (32,19,13,13) & 0.98 (2) & 0.88 (1) & 0.97 (1) & 0.93 (1) \\
e8a3eb4c & 3833,12319,12319,15595 & 4.68 (0,0,0,0) & 10.55 (0) & 10.29 (0) & 12.28 (0) & 10.14 (0) \\
eeee44ce & 183,531,531,377 & 0.74 (25,15,10,10) & 1.09 (2) & 1.06 (1) & 1.18 (1) & 1.09 (1) \\
\hline
 & Total time & 64.92 & 137.10 & 129.97 & 171.84 & 132.41 \\
\hline
\end{tabular}
\end{center}
\caption{Running time (s) and number of unalignments on the MISC-2010 Debian problem instances}\label{fig:experiments}
\end{figure}

\begin{figure}[t]
\begin{center}\scriptsize
\begin{tabular}{lllrrrr}
MISC problem id & size (\#srcs,\#vs,\#pkgs,\#pairs) & removed & packages & pairs & version changes & clusters\\
\hline
1aabfc32 & 1224,4457,4459,10304 & 2.03 (217,262,69,67) & 4.17 (0) & 6.82 (0) & 3.36 (0) & 3.04 (0) \\
1dcce248 & 3795,12207,12207,15450 & 4.90 (0,0,0,0) & 6.77 (0) & 6.36 (0) & 7.50 (0) & 6.22 (0) \\
218091ce & 3795,12207,12207,15450 & 4.51 (0,0,0,0) & 8.99 (0) & 8.75 (0) & 10.09 (0) & 8.36 (0) \\
26f3d4cc & 1130,3886,3888,7918 & 2.43 (134,111,44,43) & 4.04 (0) & 6.69 (0) & 3.63 (0) & 2.93 (0) \\
27000e82 & 1403,6088,6095,24381 & 3.12 (360,570,69,58) & 6.87 (0) & 14.91 (0) & 4.16 (0) & 4.39 (0) \\
2f690324 & 3796,12192,12192,15423 & 4.79 (0,0,0,0) & 7.53 (0) & 7.56 (0) & 9.49 (0) & 7.62 (0) \\
3e4f8550 & 341,1163,1163,1547 & 1.08 (0,0,0,0) & 1.18 (0) & 1.20 (0) & 1.29 (0) & 1.22 (0) \\
4a69cf16 & 1400,6079,6086,24445 & 2.81 (337,682,68,57) & 7.00 (0) & 14.69 (0) & 4.14 (0) & 4.42 (0) \\
4e539b28 & 1130,3886,3888,7918 & 2.37 (134,111,44,43) & 3.96 (0) & 6.67 (0) & 3.55 (0) & 2.96 (0) \\
4ede8d96 & 908,3660,3662,13895 & 1.78 (45,27,18,18) & 3.74 (0) & 7.87 (0) & 2.84 (0) & 2.95 (0) \\
5698a62c & 1400,6079,6086,24445 & 2.76 (337,682,68,57) & 6.97 (0) & 14.53 (0) & 4.14 (0) & 4.40 (0) \\
56ae4afa & 341,1163,1163,1547 & 1.06 (0,0,0,0) & 1.09 (0) & 1.10 (0) & 1.10 (0) & 1.13 (0) \\
56e31304 & 908,3660,3662,13895 & 1.78 (45,27,18,18) & 3.70 (0) & 7.83 (0) & 2.80 (0) & 2.85 (0) \\
58a4a468 & 468,1584,1584,2080 & 1.09 (0,0,0,0) & 1.16 (0) & 1.15 (0) & 1.20 (0) & 1.14 (0) \\
688250e8 & 3795,12207,12207,15450 & 6.55 (0,0,0,0) & 9.04 (0) & 8.14 (0) & 10.51 (0) & 8.55 (0) \\
6b0d1da0 & 1400,6079,6086,24445 & 2.89 (348,679,75,60) & 7.01 (0) & 14.74 (0) & 4.71 (0) & 4.53 (0) \\
7266f636 & 3802,12259,12259,15605 & 4.87 (0,0,0,0) & 6.81 (0) & 6.34 (0) & 7.45 (0) & 6.38 (0) \\
7e7e0b16 & 468,1584,1584,2080 & 1.12 (0,0,0,0) & 1.17 (0) & 1.15 (0) & 1.21 (0) & 1.14 (0) \\
8ad21cec & 3802,12259,12259,15605 & 4.82 (0,0,0,0) & 6.76 (0) & 6.42 (0) & 7.48 (0) & 6.22 (0) \\
978532fa & 1400,6079,6086,24445 & 2.77 (348,679,75,60) & 7.06 (0) & 14.80 (0) & 4.70 (0) & 4.43 (0) \\
9bb87ab4 & 3795,12207,12207,15450 & 4.54 (0,0,0,0) & 8.94 (0) & 8.79 (0) & 10.14 (0) & 8.52 (0) \\
d0cc7514 & 1400,6079,6086,24445 & 2.79 (337,682,68,57) & 6.96 (0) & 14.69 (0) & 4.11 (0) & 4.42 (0) \\
d1583bd8 & 1130,3886,3888,7918 & 2.42 (134,111,44,43) & 3.98 (0) & 6.67 (0) & 3.58 (0) & 2.85 (0) \\
dd08e73e & 1130,3886,3888,7918 & 2.40 (134,111,44,43) & 4.02 (0) & 6.73 (0) & 3.61 (0) & 2.92 (0) \\
e69a0e36 & 1426,5889,5891,20929 & 2.92 (274,265,69,60) & 5.85 (0) & 12.44 (0) & 4.44 (0) & 4.13 (0) \\
e8a3eb4c & 3795,12207,12207,15450 & 4.44 (0,0,0,0) & 8.94 (0) & 8.70 (0) & 10.20 (0) & 8.43 (0) \\
ff4a1d84 & 1224,4457,4459,10304 & 1.96 (217,262,69,67) & 4.13 (0) & 6.83 (0) & 3.46 (0) & 3.05 (0) \\
\hline
 & Total time & 81.00 & 147.84 & 222.57 & 134.89 & 119.20 \\
\hline
\end{tabular}
\end{center}
\caption{Running time (s) and number of unalignment on the Misc Live (4th run) Debian problem instances}\label{fig:experiments2}
\end{figure}

These two sets of results show a strong relationship between the structure of 
the problem, the chosen unalignment measure and the time required to solve the problem.
However, these results seems to indicate
that the version change alignment criterion offers a good trade off between discriminating power and running time.

\section{Discussion}

Aligning components in a software installation is an important issue; we have shown that
it is possible to capture this property in several ways, according to the discriminating
power one looks for, and that a state of the art MIP solver such as CPLEX has a running time
on realistic use cases that is acceptable. \\

An important question is whether a similar performance can be attained using different
solving approaches, like PBO, MaxSat or Answer Set Programming, which are present in the
MISC competition. We propose that the different measures of unalignment introduced here be
incorporated in future MISC competitions, and that component installers offer them to the users.\\

For future work, it would be interesting to allow the users to fine-tune the subset of source packages
on which the alignment is required, by introducing a more general criterion 
\textbf{unaligned(clusters:{v1,...,vn})}, that evaluates unalignment only on the clusters
for $v1,...,vn$: this does not present significant technical difficulties and can be
done by generating the constraints only for the specified source clusters.\\

Alignment being only a restricted definition of a more general
synchronization criterion, it may be equally important to synchronize
some packages that are not built from the same sources, but are
closely related. Such synchronization relations between packages could
be expressed by extending metadata.




\clearpage

\appendix
\section{Encoding unalignment for SAT}
It is possible to write natural encodings of the different criteria for SAT; we present here
the ones for the packages and package pairs criteria.

\paragraph{packages} The definition can be encoded as

\[
 unaligned_p \iff p\wedge\left(\bigvee_{ S(q_i) = S(p)  \wedge  V(q_i) \ne V(q p) } q_i\right)
\]

For minimizing unalignment, it is enough to use the clauses coming from the dominance relation

\begin{eqnarray*}
unaligned_p \Leftarrow p\wedge\left(\bigvee_{ S(q_i) = S(p)  \wedge  V(q_i)  \ne V(q p) } q_i\right) 
  & = & unaligned_p \Leftarrow \left(\bigvee_{ S(q_i) = S(p)  \wedge  V(q_i)  \ne V(q p) } p\wedge q_i\right)\\
  & = & \bigwedge_{ S(q_i) = S(p)  \wedge  V(q_i)  \ne V(q p) }(unaligned_p \Leftarrow  p\wedge q_i)\\
  & = & \bigwedge_{ S(q_i) = S(p)  \wedge  V(q_i)  \ne V(q p) }\neg unaligned_p \vee \neg p\vee \neg q_i\qquad (3)\\
\end{eqnarray*}

\paragraph{pairs} For each package pair $(p_i,p_j)$\footnote{Take $p_i<p_j$ to avoid counting the pairs twice.} which is not aligned, build a literal $unaligned_{p_i,p_j}$
which is true iff both $p_i$ and $p_j$ are installed.

\[ unaligned_{p_i,p_j} \iff p_i \wedge p_j\]

For minimizing unalignment, it is enough to use the clauses coming from the dominance relation

\begin{eqnarray*}
\lefteqn{unaligned_{p_i,p_j} \Leftarrow p_i \wedge p_j =}\\
& = &     \neg p_i \vee \neg p_j \vee unaligned_{p_i,p_j}\qquad(3)
\end{eqnarray*}

\end{document}